\documentclass[CEJP,DVI]{cej} 
\usepackage{layout}
\usepackage{amsmath}
\usepackage{textcomp}
\usepackage{hyperref}
\usepackage{subfigure}
\title{Multi-strange baryon elliptic flow in Pb-Pb collisions 
at $\sqrt{s_{NN}} = 2.76$ TeV measured with the ALICE detector}

\articletype{Research Article}

\author{Zhongbao Yin for the ALICE Collaboration\inst{1}$^,$\inst{2}\email{zbyin@mail.ccnu.edu.cn}
       }

\institute{
     \inst{1} Institute of Particle Physics, Central China Normal University,
     430079 Wuhan, China
     \inst{2} Key Laboratory of Quark \& Lepton Physics (Huazhong Normal University), Ministry of Education,
     430079 Wuhan, China
          }

\abstract{We present the results on elliptic flow with 
multi-strange baryons produced in Pb-Pb 
collisions at $\sqrt{s_{NN}} = 2.76$ TeV. The analysis is performed 
with the ALICE detector at LHC. Multi-strange baryons are reconstructed 
via their decay topologies and the $v_2$ values are analyzed with the 
two-particle scalar product method. The $p_T$ differential $v_2$ values 
are compared to the VISH2+1 model calculation and to the 
STAR measurements at 200 GeV in Au+Au collisions. We found that 
the model describes $\Xi$ and $\Omega$ $v_2$ measurements within 
experimental uncertainties. The differential flow of $\Xi$ and $\Omega$ 
is similar to the STAR measurements at 200 GeV in Au+Au collisions.}

\keywords{multi-strange baryon \*\ elliptic flow  \*\ heavy-ion collisions}
\pacs{25.75.Dw, 25.75.Gz, 21.65.Qr}

\begin{document}
\maketitle


\section{Introduction}

In heavy-ion collisions the elliptic flow ($v_2$) 
is one of the most informative observables used to understand 
the dynamics of the collisions and the 
fundamental properties of the created quark matter~\cite{Snellings11}. 
The first measurement by ALICE at 2.76 TeV~\cite{ALICE_PRL10} shows that, 
compared to flow measured at 200 GeV, the integrated 
elliptic flow of charged particles increases by 
about 30\% but the $p_T$ differential flow doesn't change within uncertainties 
at low $p_T$. This can be explained by hydrodynamical models~\cite{Shen11} 
as being due to a larger radial flow velocity at higher energies, 
which results in 
an increase of the mean $p_T$ of charged particles and 
a more pronounced mass dependence of the elliptic flow. Indeed, measurements 
of identified particle $v_2$ 
at 2.76 TeV~\cite{Krzewicki11}
indicate a larger mass splitting between pions and anti-protons 
at low $p_T < 2$ GeV/$c$ than that seen by STAR at 200 GeV~\cite{Abelev08}. 
However, to describe the anti-proton 
flow it is required to introduce a hadronic rescattering stage to the 
hydrodynamical model calculations~\cite{Heinz11}. To estimate the relative
contributions to the elliptic flow from early deconfined partonic interactions 
and later hadronic rescatterings in the system evolution, 
it is desired to measure the elliptic flow of multi-strange baryons,
as it is argued that they have small hadronic 
cross-sections~\cite{Shor85,Hecke98} 
and therefore are expected 
to be more sensitive to the quark-gluon-plasma 
phase than to the hadronic one.

In this contribution, we present the $p_T$ differential $v_2$ measurements 
of multi-strange baryons in Pb-Pb collisions at $\sqrt{s_{NN}} = 2.76$ TeV 
measured with the ALICE detector. The data are then compared to 
hydrodynamical model calculations and to RHIC measurements at 
200 GeV~\cite{Shi11}. 

\section{Data Analysis}

This analysis is based on data taken during the 2010 Pb-Pb run. 
The standard physics selection 
criteria~\cite{ALICE_PL11} for Pb-Pb collisions were applied to select 
minimum bias events. Cuts on ZDC timing, 
SPD vertex quality and TPC track multiplicity compared to the 
global track multiplicity were enforced 
at event selection level. 
Only events with a reconstructed vertex at $|v_z| < 7$ cm along 
the beam axis were used. In total, about 8 million 
selected events were used for this analysis. The event centrality 
was determined based on the sum of the amplitudes measured 
in the VZERO detectors~\cite{ALICE_PRL105}. 
Multi-strange baryons were reconstructed
via their weak decay topology: $\Xi^{-} \rightarrow \Lambda +\pi^{-}$
and $\Omega^{-} \rightarrow \Lambda + K^{-}$, with the subsequent decay of
$\Lambda\rightarrow p+\pi^{-}$. Charged tracks were reconstructed 
in the ITS and TPC detectors. Track candidates were selected in 
$|\eta| < 0.8$. Cuts on geometry, kinematics and particle identification 
on the daughter tracks via specific ionization energy loss in TPC 
were applied to reduce the 
combinatorial background. Elliptic flow was measured 
using the two-particle 
scalar product method~\cite{Adler02} 
with particles in $|\eta| < 0.5$ excluded from the flow vector 
determination. 
Only multi-strange baryon 
candidates in $|\eta| < 0.5$ were selected to analyze the elliptic flow.
The $v_2$ signal was extracted by using
$v_2 = \frac{N^{T}v_2^{T}-N^{B}v_2^{B}}{N^{S}}$,
where $v_2^{T}$ and $v_2^{B}$ are determined from the signal mass band and 
two mass side-bands of the signal region, respectively. 
$N^{T} = N^{S} + N^{B}$, where $N^{S}$ and $N^{B}$ 
are the signal and background yields in the signal mass band.

\section{Results}

\begin{figure}[h!]
\mbox{\subfigure{
\includegraphics[width=3.3 in]{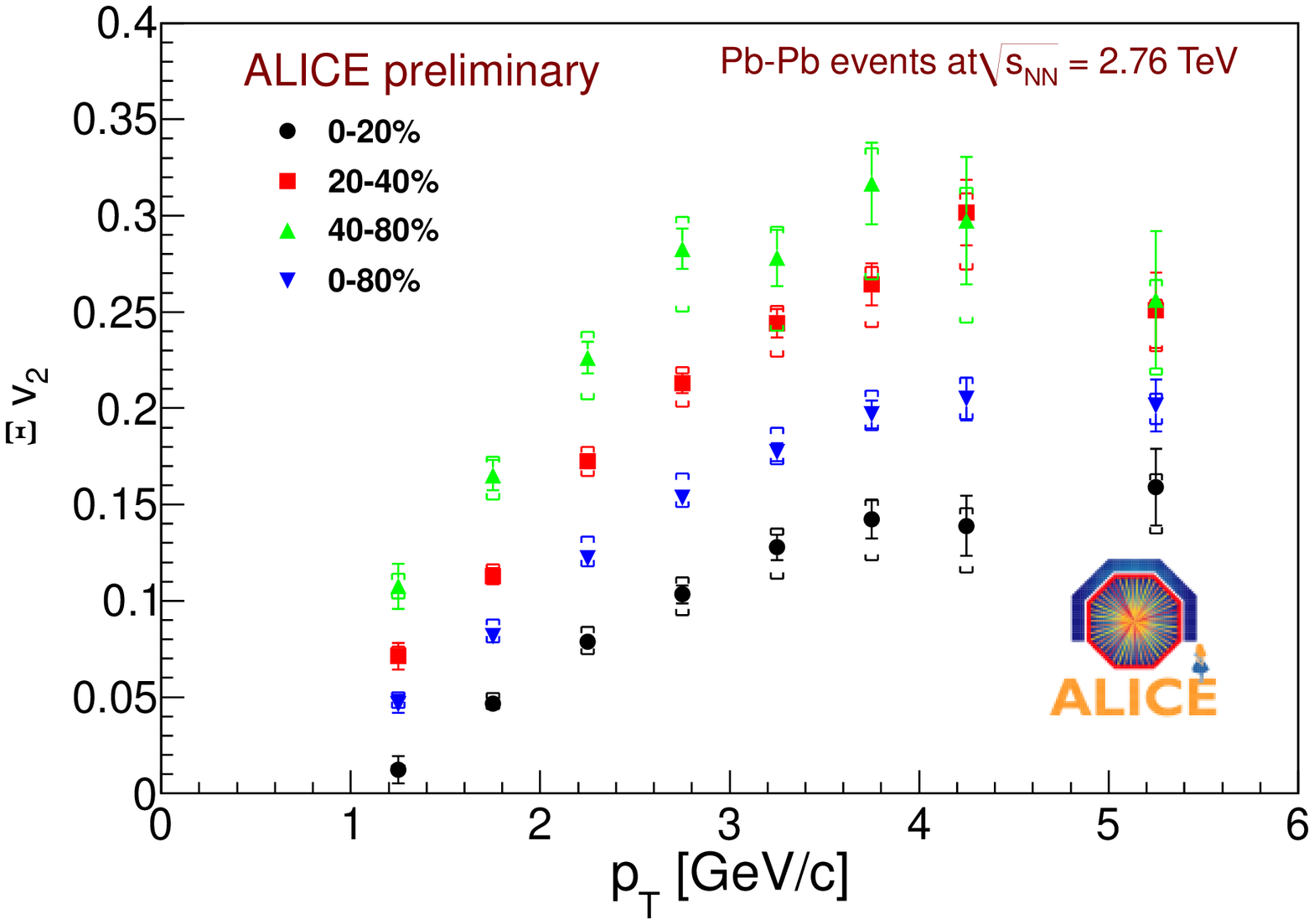}}
\quad
\subfigure{
\includegraphics[width=3.3 in]{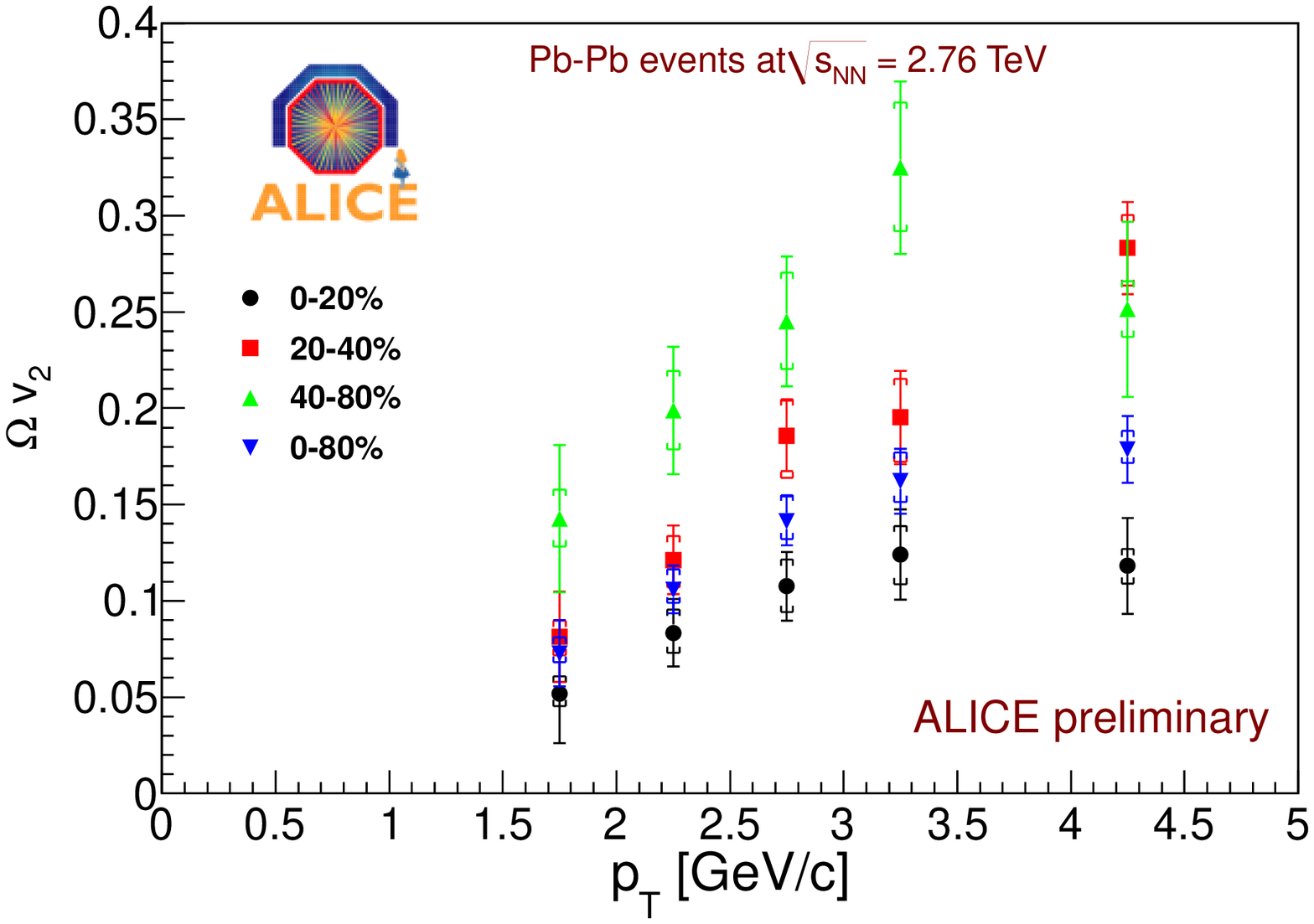}}}
\caption{$v_2(p_T)$ of $\Xi^{-}+\bar{\Xi}^{+}$ (left) and $\Omega^{-}+\bar{\Omega}^{+}$ (right) for 0-20\%, 20-40\%, 40-80\% and 0-80\% centralities in 2.76 TeV Pb-Pb collisions. The error bars show the statistical uncertainties. The brackets indicate the systematic errors.
\label{fig:Xi}}
\end{figure}

\begin{figure}[h!]
\centering
\begin{minipage}{0.495\textwidth}
\centering
\includegraphics[width=3.2 in]{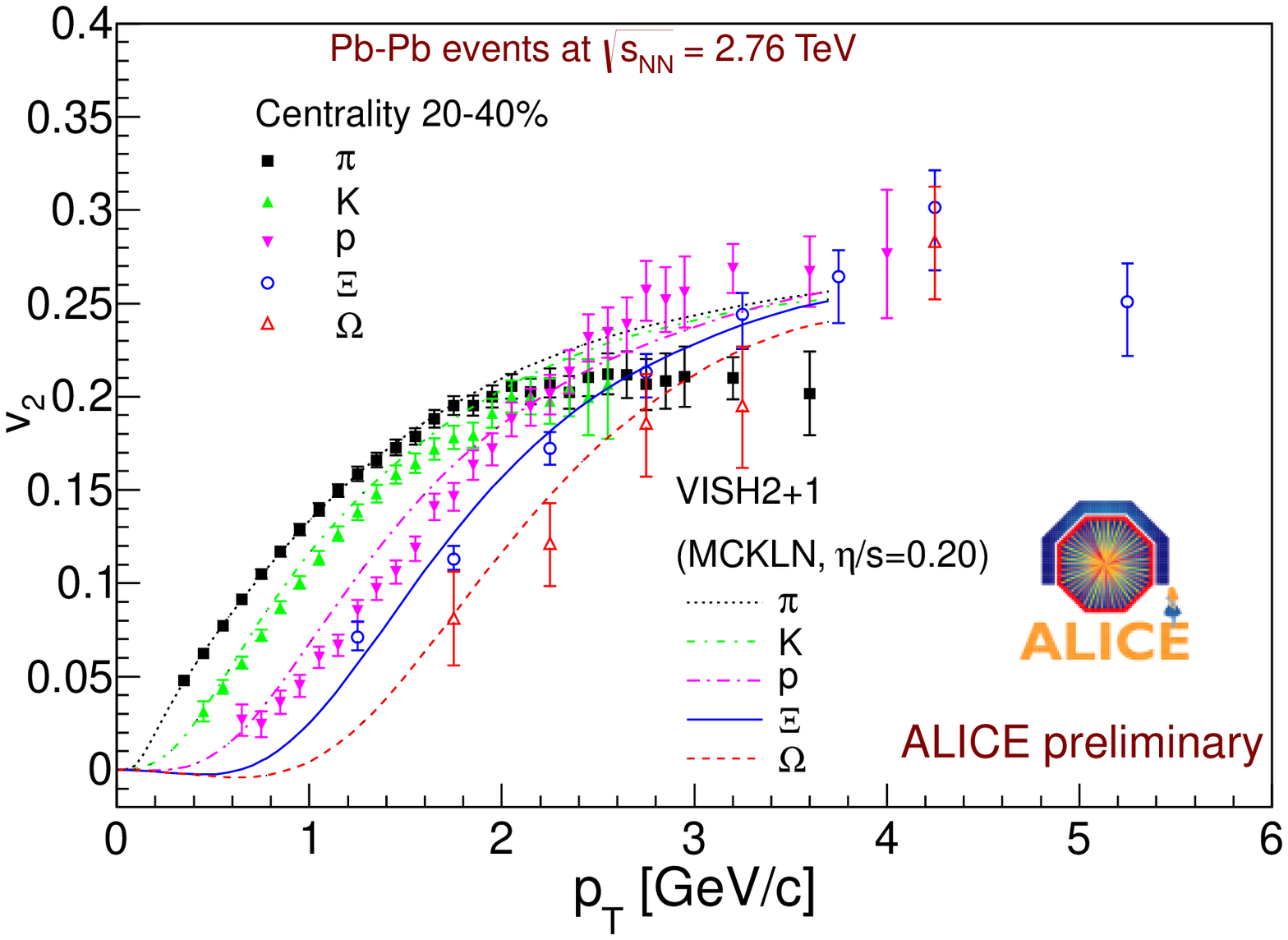}
\caption{$v_2(p_T)$ of identified particles in 20-40\% centrality Pb-Pb collisions at 2.76 TeV compared to VISH2+1 hydrodynamical model calculations.
\label{fig:VISH2p1}}
\end{minipage}
\begin{minipage}{0.495\textwidth}
\centering
\includegraphics[width=3.3 in]{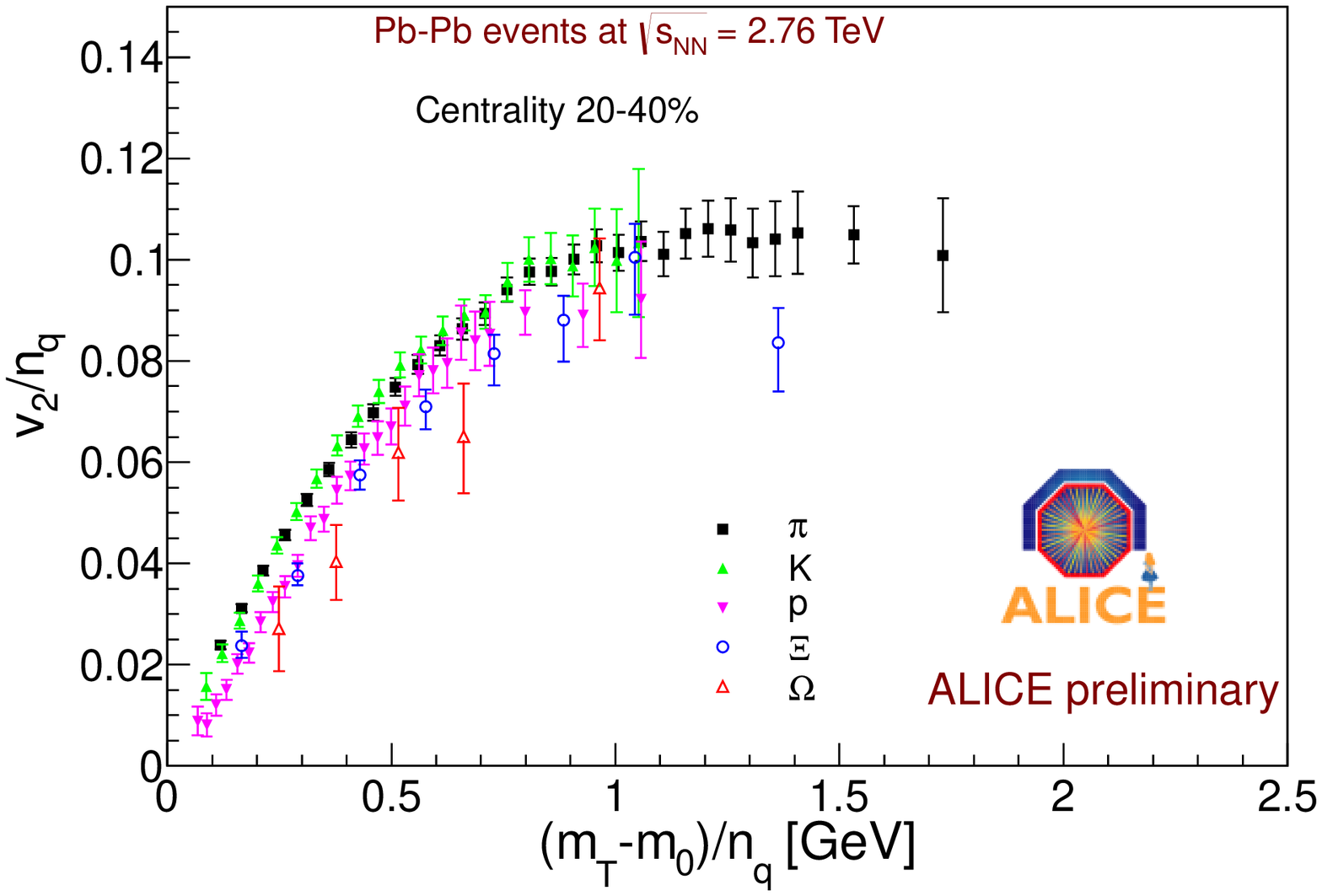}
\caption{$v_2/n_q$ versus $(m_T-m_0)/n_q$ for identified particles in 20-40\%
centrality Pb-Pb collisions at 2.76 TeV.\label{fig:KEtScaling}}
\end{minipage}
\end{figure}

Figure~\ref{fig:Xi} shows $v_2(p_T)$ of $\Xi^{-}+\bar{\Xi}^{+}$ (left) 
and $\Omega^{-}+\bar{\Omega}^{+}$ (right) for 0-20\%, 20-40\%, 40-80\% 
and 0-80\% centrality classes in 2.76 TeV Pb-Pb collisions. 
The error bars show the statistical uncertainties. The brackets 
indicate the systematic errors. Figure~\ref{fig:VISH2p1} shows elliptic flow 
of identified particles measured for 20-40\% central collisions compared with 
VISH2+1 hydrodynamical model calculations~\cite{Heinz11}. 
The error bars in this figure 
(and also in the following ones) indicate systematic and statistical 
uncertainties added in quadrature. As can be seen, the VISH2+1 model 
describes the measured $v_2$ for pions and kaons, but deviates up 
to 20\% for anti-protons. Within experimental uncertainties the model 
calculations are compatible with the measured 
$\Xi$ and $\Omega$ elliptic flow. 

To test whether the transverse kinetic energy scaling observed at RHIC 
energies~\cite{Adare07} holds at LHC energies, we plot in 
Fig.~\ref{fig:KEtScaling} $v_2/n_q$ versus $(m_T-m_0)/n_q$ for identified 
particles in 20-40\% centrality Pb-Pb collisions at 2.76 TeV. 
The transverse kinetic energy $n_q$ scaling of elliptic flow is not observed, 
but within errors mesons and baryons separately seem to scale.

\begin{figure}[hbt!]
\mbox{\subfigure{
\includegraphics[width=3.3 in]{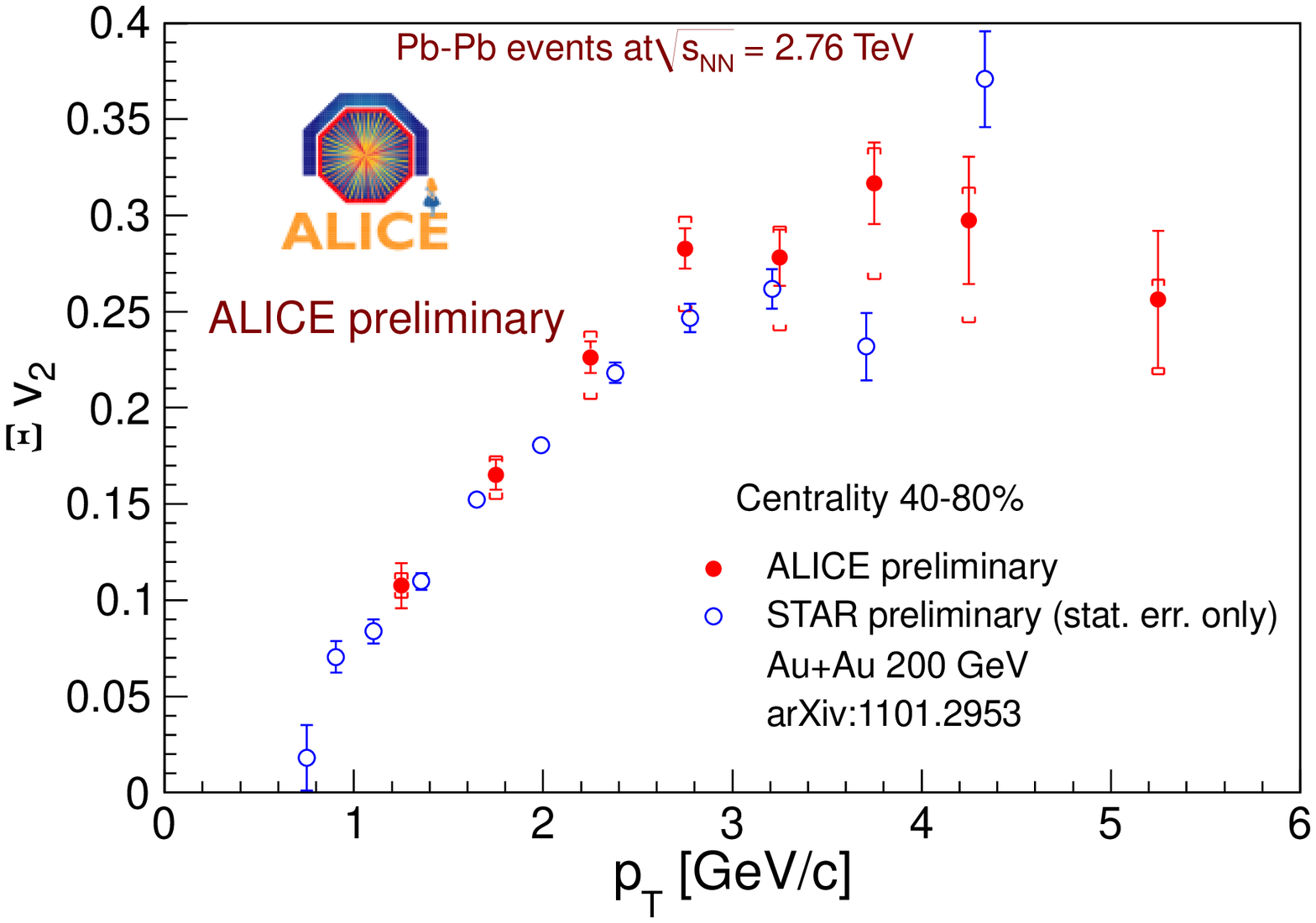}}
\quad
\subfigure{
\includegraphics[width=3.3 in]{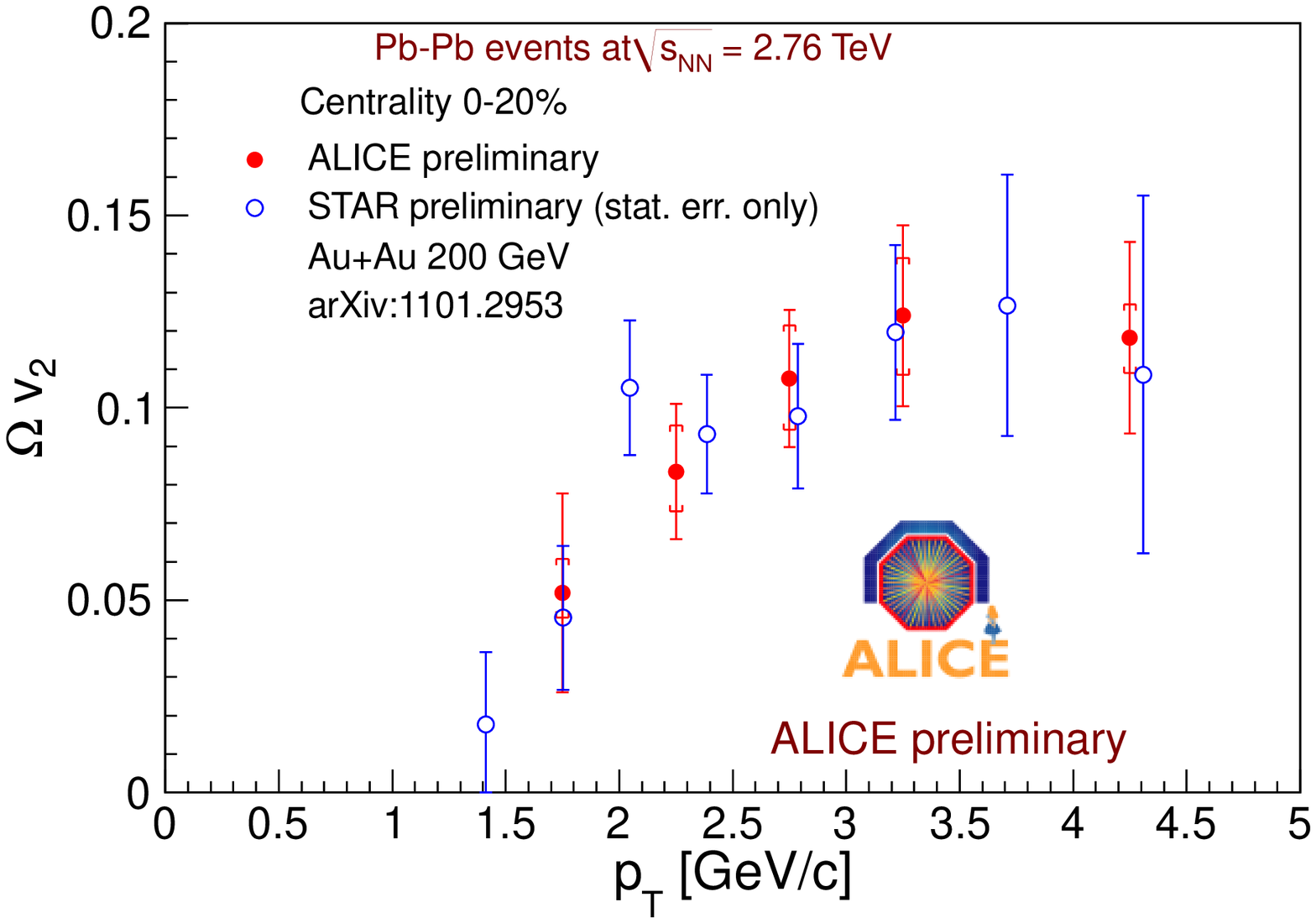}}}
\caption{Comparison of $\Xi^{-}+\bar{\Xi}^{+}$ $v_2$ (left) for 40-80\% centrality and $\Omega^{-}+\bar{\Omega}^{+}$ $v_2$ (right) for 0-20\% 
centrality in 2.76 TeV Pb-Pb collisions to the STAR measurements in 200 GeV 
Au-Au collisions. 
\label{fig:Comparison2STAR}}
\end{figure}

Figure~\ref{fig:Comparison2STAR} shows a comparison 
of $\Xi^{-}+\bar{\Xi}^{+}$ $v_2$ (left) for 40-80\% centrality 
and $\Omega^{-}+\bar{\Omega}^{+}$ $v_2$ (right) for 0-20\%
centrality in 2.76 TeV Pb-Pb collisions to results from 
200 GeV Au-Au collisions at the same centralities 
measured by the STAR experiment~\cite{Shi11}. 
As can be seen, the differential flow 
does not change within uncertainties. 

\section{Summary}

We presented the $p_T$ differential elliptic flow of multi-strange baryons in 
Pb-Pb collisions at $\sqrt{s_{NN}} = 2.76$ TeV measured 
with the ALICE detector at the LHC.
We showed the $p_T$ differential elliptic flow of identified particles 
for 20-40\% centrality and compared it to the VISH2$+$1 model 
calculations. The viscous-hydro prediction describes $\Xi$ 
and $\Omega$ $v_2$ measurements within experimental uncertainties.
We also showed that the number of constituent quark transverse kinetic 
energy scaling does not work as well as at RHIC. Mesons and baryons 
seem to follow two different 
trends when the $v_2/n_q$ is plotted as a function of $(m_T-m_0)/n_q$.
In addition, we observed that the differential flow 
of $\Xi$ for 40-80\% centrality and $\Omega$ for 0-20\% centrality 
is similar to the STAR measurements at 200 GeV in Au-Au collisions.

\section*{Acknowledgments}
We thank the ALICE collaboration and the ALICE funding agencies (the same 
as the acknowledgments in~\cite{ALICE_PL11}). This work is supported partly by 
the NSFC under grants No. 10975061, 10875051 and 11020101060.

\end{document}